\RequirePackage{lineno}
\documentclass{emulateapj}

\usepackage{amsmath,natbib,graphicx}
\usepackage{epsf}
\usepackage{fancyvrb}
\usepackage{epstopdf}
\usepackage{epsfig}
\usepackage{color}
\DeclareGraphicsExtensions{.jpg,.pdf,.png,.eps,.ps}

\begin{document}


\VerbatimFootnotes
\title{A Low-order Model of Water Vapor, Clouds, and 
Thermal Emission for Tidally Locked Terrestrial Planets}

\shorttitle{A Low-order Climate Model for Tidally Locked Planets}

\author{Jun Yang and Dorian S. Abbot}
\affil{Department of the Geophysical Sciences, University of Chicago, 
5734 South Ellis Avenue, Chicago, IL 60637, USA}
\email{Correspondence: junyang28@uchicago.edu}

\shortauthors{Yang and Abbot 2014 (in press)}

\begin{abstract}
  In the spirit of minimal modeling of complex systems, we develop an
  idealized two-column model to investigate the climate of tidally
  locked terrestrial planets with Earth-like atmospheres 
  in the habitable zone of M-dwarf stars.
  The model is able to approximate the fundamental features of the
  climate obtained from three-dimensional (3D) atmospheric general
  circulation model (GCM) simulations. One important reason for the
  two-column model's success is that it reproduces the high cloud
  albedo of the GCM simulations, which reduces the planet's
  temperature and delays the onset of a runaway greenhouse state. The
  two-column model also clearly illustrates a secondary mechanism
  for determining the climate: the nightside acts as a ``radiator fin''
  through which infrared energy can be lost to space easily. This
  radiator fin is maintained by a temperature inversion and dry air on
  the nightside, and plays a similar role to the subtropics on modern
  Earth.  Since 1D radiative-convective models cannot capture the
  effects of the cloud albedo and radiator fin, they are
  systematically biased towards a narrower habitable zone. We also
  show that cloud parameters are most important for determining the
  day--night thermal emission contrast in the two-column model, which
  decreases and eventually reverses as the stellar flux increases.
  This reversal is important because it could be detected by future
  extrasolar planet characterization missions, which would suggest
  that the planet has Earth-like water clouds and is potentially
  habitable.
  
\end{abstract}


\section{Introduction}

Extrasolar planets in tidally locked orbital configurations around
low-mass and relatively cool M-dwarf stars are the prime targets of
several ongoing and proposed planet search programs
\citep{Tarteretal2007}.  In advance of direct observations of these
planets, numerical models, including both single column radiative--convective models
\citep[e.g.,][]{Kastingetal1993,Wordsworthetal2010, HuandDing2011, Kopparapuetal2013}
and three-dimensional atmospheric general circulation models
\citep[GCMs,][hereafter, YCA13]{Joshietal1997, Joshi2003,
  MerlisandSchneider2010, HengandVogt2011, Pierrehumbert2011,
  Edsonetal2011, Wordsworthetal2011, Leconteetal2013, Shieldsetal2013,
  Yangetal2013} have been employed to investigate their potential
climates.  Radiative--convective models employ sophisticated radiative
schemes, but neglect atmospheric dynamics. GCMs calculate
atmospheric dynamics in detail, and use radiative schemes of varying
levels of complexity.

In addition to numerically intensive modeling, idealized
models can help make clear essential physical mechanisms in complex
systems such as the climates of tidally locked planets. Low-order
models are relatively easy to analyze and determine the dominant
mechanisms, since mechanisms can be easily added, removed, or
changed. They can be compared with more complex models to aid the
interpretation of complex model results, and they help to address the
critical question of what is the minimum physics necessary to
understand a particular problem. 

In this study, we develop a two-column model of the atmospheres of
Earth-like tidally locked planets in order to better understand the
following two questions: (1) What determines the surface temperature?
and (2) What determines the thermal infrared emission contrast between
dayside and nightside? These questions are critical because the
surface temperature determines whether liquid water can be maintained
on the surface, which determines habitability in the traditional
sense, and the thermal emission contrast will be one of the first
observational signals that will be used to characterize tidally locked
terrestrial planets \citep[e.g.,][]{Knutsonetal2007}.

The model divides the atmosphere into two columns with one
representing the dayside and the other representing the
nightside (Fig.~1). This is the most basic possible
simplification that allows horizontal heterogeneity, and it is
justified by the following facts: (1) Only the dayside receives
stellar flux from the parent star, whereas the nightside is heated by
atmospheric and ocean heat import from the dayside. The temperatures
of the two sides are therefore determined by distinct physical
processes. (2) GCM simulations have found that on the nightside
horizontal surface and air temperature gradients are extremely small
and atmospheric descent occurs throughout the nightside
\citep{Joshietal1997, MerlisandSchneider2010}; therefore, the
nightside atmosphere can be approximated as a single column. (3) On
the dayside, the surface temperature is homogeneous in the vicinity of
the substellar point in GCM simulations, and more broadly much of the
dayside is characterized by robust moist convection and mean ascent
\citep[YCA13;][]{MerlisandSchneider2010}, so that it can be roughly
treated as a single column. (4) The dayside has a moist atmosphere
with a robust hydrological cycle whereas the nightside atmosphere is
extremely dry \citep{MerlisandSchneider2010, Edsonetal2011}. (5)
Future observations of horizontal variations in the climate on tidally
locked terrestrial planets are likely to be coarse, so it is
reasonable to develop a model with minimal horizontal
resolution.

In our model, the dayside column is thought of as a moist region with
deep convection, mean ascent, and a warm surface.  The air temperature
follows a saturation moist adiabatic temperature profile. In this way
the dayside is similar to the warm pool of the tropical Pacific Ocean
on Earth, where the temperature profile is close to a moist adiabat
\citep{XuandEmanuel1989}. The nightside column has mean descent with
dry air and a cold surface, which is analogous in some ways the
subtropics of Earth.  The implicit atmospheric circulation connecting
the dayside and the nightside in the model approximates a global-scale
Walker circulation driven by the stellar energy contrast, as obtained
in GCM simulations (see the review of \cite{Showmanetal2013}). To
simulate the effect of clouds, we employ a simple convective cloud
scheme and tune one free parameter of the scheme to produce a similar
cloud albedo to that obtained in the GCM simulations of YCA13.  At a
given stellar flux and a specified ocean heat transport, the model
computes dayside and nightside surface temperature, dayside and
nightside free-tropospheric temperature, dayside convective heat flux,
and atmospheric heat transport from the dayside to the nightside.

\begin{figure}[h!]
\vspace{-15mm}
\begin{center}
\includegraphics[angle=0, width=27pc]{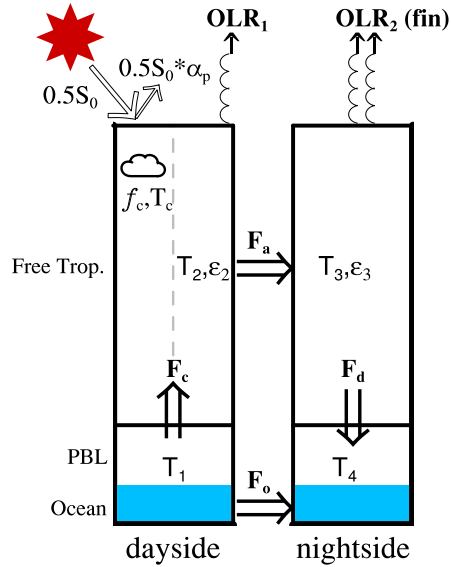}
\end{center}
\vspace{-8mm}
\caption{Schematic representation of the two-column model for the
  climate of tidally locked terrestrial planets. The dayside column
  consists of a cloudy part and a clear-sky part. S$_0$: stellar flux
  at substellar point; $\alpha_p$: planetary albedo; $OLR$: outgoing
  longwave radiation (infrared emission to space); $f_c$: effective
  cloud fraction; $T_c$: cloud emission temperature;
  $\varepsilon_2,_3$: atmospheric emissivity; PBL: planetary boundary
  layer; $F_c$: convective heat flux; $F_a$: atmospheric heat
  transport from dayside to nightside; $F_o$: ocean heat transport;
  and $F_d$: the fraction of atmospheric heat transport from the
  dayside to the nightside deposited in the nightside boundary layer
  due to adiabatic heating.}
\label{fig1}
\end{figure}


Similar types of models have been employed to investigate the climate
of the tropics on modern Earth. Such models have been found to be useful
for identifying the mechanisms that regulate the surface temperature
of the tropical Pacific Ocean, including surface evaporation
\citep{HartmannandMichelsen1993}, convective clouds
\citep{RamanathanandCollins1991}, stratus low clouds
\citep{Larsonetal1999}, water vapor greenhouse effect
\citep{Pierrehumbert1995}, horizontal atmospheric heat transport
\citep{HartmannandMichelsen1993, Pierrehumbert1995}, and ocean
dynamics \citep[e.g.,][]{ClementandSeager1999}. Because of the extreme
simplifications of these column models, their applicability to direct
comparison with detailed and spatially dense observations is somewhat
limited; however, they have significantly improved our understanding
of tropical climate by helping to identify the essential processes
that govern the  system.

\begin{figure*}[]
\begin{center}
\vspace{-40mm}
\begin{center}
\includegraphics[angle=0, width=34pc]{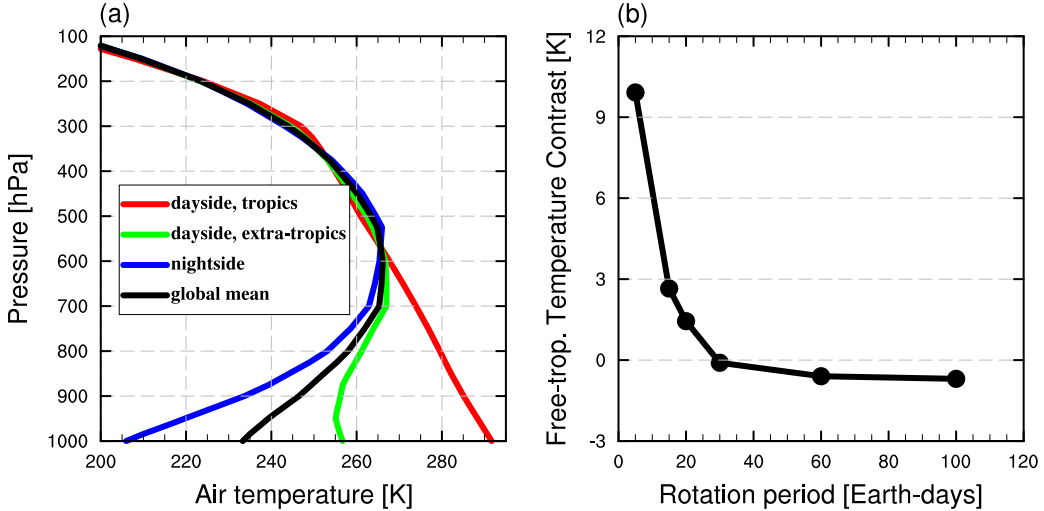}
\end{center}
\vspace{-35mm}
\caption{GCM output that supports the weak-temperature-gradient
  approximation for tidally locked planets. (a) Vertical temperature
  profiles for a tidally locked configuration with a rotation period
  of 60 Earth-days. Red line: for the tropics
  (30$^{\circ}$S--30$^{\circ}$N) of the dayside; green line: for the
  extra-tropics of the dayside; blue line: for the entire nightside;
  and black line: global mean.  (b) Free-tropospheric temperature
  contrast between dayside and nightside (mean value between the
  pressure levels of 100 and 600 $hPa$) as a function of rotation
  period. The stellar flux is 1200\,W\,m$^{-2}$ in these
  simulations.}
\label{fig2}
\end{center}
\end{figure*}

We tune our two-column model to a GCM, which fully resolves
atmospheric circulation, radiative transfer, the hydrological cycle,
and clouds.  We use an Earth
GCM, the Community Atmosphere Model version 3 (CAM3;
\cite{Collinsetal2004}), developed by the National Center for
Atmospheric Research. The GCM is coupled to a mixed layer ocean
with a uniform depth of 50 m. We have modified the GCM to
simulate the climate of tidally locked planets around M-stars (for
details, see YCA13).  The default planetary radius, gravity, and
rotation period are set to two times Earth's, 13.7\,m\,s$^{-2}$, and 60
Earth-days, respectively.  The atmosphere is composed of N$_2$ and
H$_2$O with a surface pressure of 1 bar, and other greenhouse gases
such as CO$_2$, CH$_4$, N$_2$O, and O$_3$ are set to zero. Three groups of
experiments with different stellar fluxes (from 1000 to
2400~W\,m$^{-2}$), ocean heat transport (from 0 to 55~W\,m$^{-2}$),
and rotation periods (from 5 to 100~Earth-days) were performed, and
the results are used here for comparisons with the two-column model.
The GCM and the two-column model share the same planetary and
atmospheric parameters, such as planetary gravity and specific heat of air.

The goal of this study is to increase understanding of the key
controls on surface temperature and thermal emission flux of tidally
locked Earth-like planets. The basic physical processes that we build
into the model are outlined in Section 2. The model is developed in
Section 3. Section 4 presents the behavior of the two-column model,
comparisons of model results with GCM simulations, and sensitivity
analyses of the model.  In Section 5, the critical stellar flux at
which the day--night thermal emission contrast becomes negative is
addressed. This contrast is essential for the interpretation of phase
curves of terrestrial planets that will be measured in the near
future.  Section 6 summarizes the main findings of this study.


\section{Essential Physical Processes}

The orbital distance of tidally locked planets in the habitable zone
of M stars should be in the range of $\approx$0.02--0.2 AU (1 AU is
the distance between Earth and Sun) \citep[YCA13;][]{Kastingetal1993,
  Kopparapuetal2013}, corresponding to rotation periods of tens of
Earth-days. The extreme variation in stellar radiation around the
planet and the slow rotation rate (corresponding to a weak Coriolis
force) imply that in many ways the climate of tidally locked planets
will be quite different from that of Earth. Despite this, we should
still be able to use basic principles of climate dynamics discovered
by studying Earth to better understand the climate of these planets.
In this section, we outline the essential physical mechanisms that we
believe govern the climate of tidally locked terrestrial planets,
including the weak-temperature-gradient approximation, a stabilizing
cloud feedback, and the fixed anvil temperature hypothesis. We will
also discuss the factors determining thermal emission to space.

\subsection{The Weak-temperature-gradient Approximation}

GCM simulations suggest that the free-tropospheric temperature of
tidally locked habitable planets should be quite horizontally uniform
(Fig.~2a) due to the weak Coriolis force. This property is similar to
the condition of the tropics on Earth
\citep{Pierrehumbert1995,Sobeletal2001} where the Coriolis force is
also weak, which allows the horizontal temperature gradients of the free
troposphere to be approximated as zero. This is referred to as the
weak-temperature-gradient (WTG) approximation. It is important
  to note that the WTG approximation only applies to the atmosphere
  above the planetary boundary layer. In the planetary boundary layer,
  where frictional forces become important in the momentum balance,
  horizontal temperature gradients can remain large, so that the
  nightside surface can become much colder than the dayside
  surface. On a tidally locked planet, where strong inversions develop
  on the nightside and at high latitudes on the dayside, a stably
  stratified planetary boundary layer can extend up to $\approx$600~hPa 
  (see Fig.~2a). Nevertheless, a model approximating the atmosphere
  with just two levels, the boundary layer (and surface) and the free
  troposphere, can yield a faithful representation of the longitudinal
  variation in infrared emission to space \citep{MillsandAbbot2013}.

\begin{figure*}[]
\begin{center}
\vspace{-35mm}
\begin{center}
\includegraphics[angle=0, width=34pc]{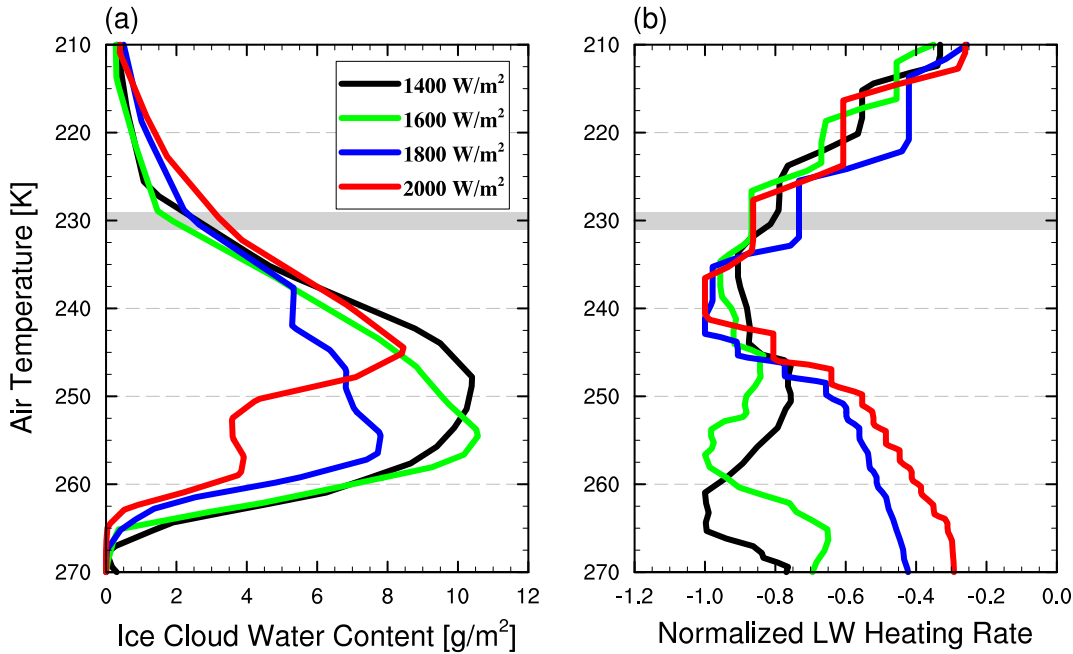}
\end{center}
\vspace{-25mm}
\caption{GCM output justifying the fixed anvil temperature
  hypothesis for tidally locked planets. (a) Dayside ice cloud water content, and (b) 
  normalized clear-sky longwave heating rate, as a function of air temperature. 
  The horizontal gray bold line roughly shows the temperature at the top 
  of ice clouds where a sharp decrease in the clear-sky longwave heating rate occurs. 
  Note that the pressure at which $T$\,=\,230~$K$ occurs is $\approx$150, 100, 
  40 and 12 $hPa$ for the stellar fluxes of 1400, 1600, 1800, and 2000~W\,m$^{-2}$, respectively.}
\label{fig3}
\end{center}
\end{figure*}

The weak temperature gradient in the free troposphere can be brought
about by a variety of physical processes. Temperature gradients set up
gravity waves that quickly diminish them. Additionally, the
atmospheric circulation of tidally locked planets is characterized by
a global-scale Walker circulation: strong updrafts near the substellar
point, divergence at high altitudes, broad downwelling at the entire
nightside, and convergence flows at low altitudes, returning to the
substellar region \citep{Showmanetal2013}.  Moreover, the zonal-mean
zonal wind velocity is dominated by a weak eastward superrotation along
the equator.  This equatorial superrotation results from equatorward
zonal momentum transport by Rossby waves induced by strong zonal
variations in the stellar radiation
\citep[e.g.,][]{ShowmanandPolvani2011}. All of these processes are
important for transporting heat from dayside to nightside and
maintaining the weak temperature gradient in the troposphere.

The WTG approximation is not valid if the timescale for atmospheric
heat transport is longer than the radiative timescale. Two ways for
this to happen are if the planet has a small orbital (and rotational)
period or has an extremely hot and/or thin atmosphere
\citep{Perez-BeckerandShowman2013}. For our sample GCM simulations
the largest horizontal temperature differences in the free troposphere
around the planet are only $\approx$10~K when the orbital period is 5
Earth-days (Fig.~2b). This is consistent with previous work indicating
that the WTG approximation does a good job of approximating the
temperature profile of a tidally locked planet even if the rotation
period is only 1 Earth-day 
\citep{MerlisandSchneider2010,MillsandAbbot2013}, and suggests that
the WTG approximation may be more broadly applicable than one might initially
suspect.


\subsection{The Stabilizing Cloud Feedback}

Clouds contribute most of the planetary albedo of Earth
\citep{DonohoeandBattisti2011}. Consistent with the global-scale
Walker circulation and ascent at the substellar region, most
parts of the dayside of tidally locked Earth-like planets would be
covered by water clouds \citep[YCA13;][]{Edsonetal2011}, which would
greatly increase the planetary albedo and
dramatically cool the planet. The cloud albedo effect becomes
stronger as the planet is moved closer to its parent star because
higher stellar insolation drives stronger convection, producing more
clouds and increasing their optical depth.  This stabilizing cloud
feedback significantly delays the onset of a runaway greenhouse state
as the stellar flux increases and therefore shifts the inner edge
of the habitable zone substantially toward its star (YCA13).

In the two-column model, we parametrize convection as a vertical
mixing of moist static energy (MSE) between the boundary layer and the
free troposphere. Convection commences when the boundary-layer MSE
exceeds the free-tropospheric saturation MSE \citep{Raymond1995}. The
cloud fraction and cloud albedo are assumed to be proportional to the
logarithm of the convective heat flux, as suggested by observations of
tropical convective clouds on Earth \citep{Slingo1980}. The
logarithm formula connects the optical properties of clouds to the
strength of convection, and enables the model to capture the effect of
the stabilizing cloud feedback.


\subsection{The Fixed Anvil Temperature Hypothesis}


\begin{figure*}[]
\begin{center}
\vspace{-50mm}
\begin{center}
\includegraphics[angle=0, width=34pc]{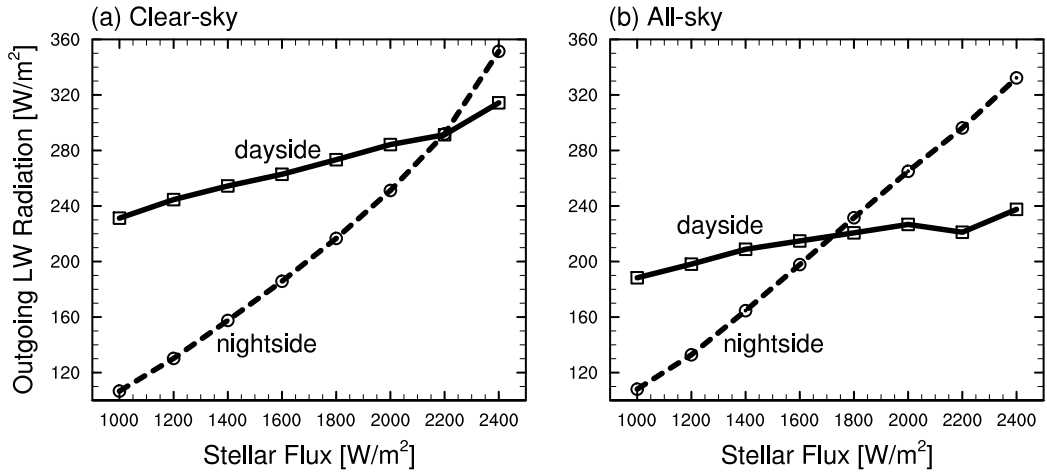}
\end{center}
\vspace{-40mm}
\caption{GCM simulated outgoing longwave radiation at the top of the 
atmosphere on the dayside (solid line) and on the nightside (dashed line) as 
a function of the stellar flux ($S_0$). (a) Clear-sky conditions without clouds, 
and (b) all-sky conditions including the effect of clouds. 
Note (a) and (b) are calculated with the same surface temperatures, 
air temperatures, and water vapor concentrations.}
\label{fig4}
\end{center}
\end{figure*}

Studies of tropical anvil clouds on Earth suggest that their emission
temperature is nearly constant as the climate changes. This is
referred to as the fixed anvil temperature (FAT) hypothesis
\citep{HartmannandLarson2002}.  Energy balance in the tropical
troposphere is primarily between convective heating by latent heat
release in regions of deep convection and radiative cooling by
longwave emission to space in clear-sky regions (i.e., the regions
without clouds) with large-scale subsidence. Because of this, the
detrainment level (where outflow occurs) of anvil clouds should be
located at the altitude where the clear-sky radiative cooling
diminishes rapidly.  The clear-sky radiative cooling rate in the upper
troposphere is primarily determined by water vapor emission. The
temperature at which the saturation water vapor pressure becomes small
enough that water vapor emission is ineffective is narrowly
constrained by local air temperature because of the
Clausius--Clapeyron relation, and does not depend on surface
temperature. The FAT hypothesis asserts that the temperature at which
anvil clouds detrain is determined by the level where a large decrease
in water vapor radiative cooling occurs. According to the FAT
hypothesis, the temperature at the top of anvil clouds should be
nearly independent of surface temperature, which exerts a strong
constraint on the strength of cloud longwave forcing.

Recent simulations with cloud-resolving models
\citep[e.g.,][]{KuangandHartmann2007} and analyses with cloud object
data \citep[e.g.,][]{Lietal2012} support the FAT hypothesis.
Furthermore, \cite{ZelinkaandHartmann2010} show that the cloud-top
temperature in GCMs with parameterized clouds also remain
approximately constant as greenhouse gas concentrations are
increased. Our GCM simulations of tidally locked planets also support
the FAT hypothesis. As shown in Fig.~3, the clear-sky cooling rate
decreases rapidly at a constant air temperature, and the cloud-top
temperature stays constant when the stellar flux is increased
massively. In the two-column model, it is therefore reasonable to
set the cloud-top temperature to a constant.


\subsection{Factors Determining Thermal Emission to Space}

A primary tool for deciphering the climate of terrestrial exoplanets
will be to examine phase variations in their infrared emission flux,
commonly referred to as outgoing longwave radiation in Earth science
($OLR$). The $OLR$ is primarily determined by surface temperature
($T_s$), air temperature ($T_a$), atmospheric specific humidity ($q$,
the mass ratio of water vapor to dry air), and clouds.  Changes in
temperature and humidity have opposing effects on $OLR$: increases in
$T_s$ and $T_a$ increase $OLR$, while increases in $q$ decrease $OLR$
due to the water vapor greenhouse effect \citep{Pierrehumbert2010}.
Clouds can also significantly reduce $OLR$ \citep{Ramanathanetal1989}.
$OLR$ can be thought of as the sum of surface upward emission which
survives absorption by water vapor, reemission from the atmosphere,
and a negative term due to cloud absorption,
 \begin{equation}
OLR = (1-\varepsilon(q)) \sigma T_s^4 + \varepsilon(q) \sigma T_a^4 - C_l,
\end{equation}
where $\varepsilon(q)$ is the clear-sky atmospheric infrared
emissivity as a function of specific humidity ($q$), $C_l$ is the
cloud longwave forcing (i.e., the change in $OLR$ in the presence of
clouds relative to the clear-sky condition), and $\sigma$ is the
Stefan-Boltzmann constant. For an optically thin clear-sky atmosphere,
$\varepsilon$ is close to zero, and $OLR$ is close to the surface
emission. For an optically thick clear-sky atmosphere, $\varepsilon$
is close to one, and $OLR$ is primarily determined by the emission
from the upper levels of the atmosphere.


For tidally locked terrestrial planets, the dayside specific
  humidity tends to be higher than the nightside specific humidity at
  all atmospheric levels. This is because the relative humidity tends
  to be higher at all levels on the dayside, and additionally the
  dayside temperature is much higher in the boundary layer. As a
result, the rate of increase of the dayside clear-sky $OLR$ with
stellar flux is much smaller than that on the nightside (Fig.~4a). The
high water vapor concentration shifts the atmospheric emission level
to higher altitudes with colder temperatures, which keeps the dayside
$OLR$ from increasing much.

The dayside is covered by high-level clouds which strongly
  reduce the dayside OLR whereas the nightside is dominated by
  low-level clouds that have little effect on the nightside OLR.
High-level clouds absorb thermal emission from the warm surface and
reemit it to space following the Stefan--Boltzmann law at the
temperature of the cloud, which is much lower than the surface
temperature, reducing $OLR$. Low-level clouds have nearly zero effect
on $OLR$ because they have temperatures close to the surface
temperature. GCM simulations show that as a result of the global
circulation pattern, with ascent on the dayside and descent on the
nightside, the dayside should be covered by both high-level and
low-level clouds while the nightside should be dominated by low-level
clouds (Fig.~1 of YCA13) that form when water vapor is trapped in the
boundary layer under the temperature inversion (see Fig.~2a). Because
of the combined effect of clouds and water vapor, the dayside all-sky
(including clouds) $OLR$ is nearly constant as the stellar flux is
varied (Fig.~4b).

In our two-column model, the clear-sky region of the atmosphere is
treated as an equivalent gray gas atmosphere characterized by an
emissivity that is a function of water vapor content
\citep{Pierrehumbert2010}.  This representation qualitatively captures
the greenhouse effect of water vapor without requiring detailed
calculations of the radiative transfer. In order to calculate the
cloud longwave forcing, the dayside column also includes a cloudy
region that emits at a constant temperature (following the FAT
hypothesis).  This scheme does a reasonable job of simulating the
cloud longwave effect.


\section{The Two-column Model}

In this section, we construct an idealized two-column model (shown
schematically in Fig.~1) that incorporates all the physical processes
addressed in Section 2 above.  The dayside column is thought of as a
region with deep convection, net ascent, and a moist atmosphere. 
The nightside column has no convection and represents a dry region of mean
descent. In each column, the upper level represents the free
troposphere and the lower level represents both the boundary layer and
the surface. The boundary layer and surface are assumed
to be tightly coupled by turbulent mixing and therefore have
approximately the same temperature. The dayside free troposphere 
has two parts, the clear-sky part and the cloudy part. 

The model is formulated by requiring energy balance in the free
troposphere and at the surface (Eqs.~(2--5)), making the WTG
approximation (Eq.~(6)), enforcing convective neutrality on the
dayside (Eq.~(7)), and assuming the FAT hypothesis is correct
(Eq.~(8)). The model equations are:
\begin{equation}
\frac{1}{2}S_0(1-\alpha_p) - F_c - F_o + (1 - f_c) \varepsilon_2 \sigma T_2^4 
+ f_c \sigma T_c^4 - \sigma T_1^4  = 0,
\end{equation}
\begin{equation}
F_c - F_a + (1- f_c) \varepsilon_2 \sigma T_1^4 + f_c \sigma T_1^4 
- 2(1 - f_c) \varepsilon_2\sigma T_2^4 - 2 f_c \sigma T_c^4 = 0,
\end{equation}
\begin{equation}
F_a - F_d + \varepsilon_3 \sigma T_4^4 - 2 \varepsilon_3\sigma T_3^4 = 0,
\end{equation}
\begin{equation}
F_o + F_d + \varepsilon_3 \sigma T_3^4 - \sigma T_4^4 = 0,
\end{equation}
 \begin{equation}
T_2 - T_3 =0,
\end{equation}
 \begin{equation}
MSE_1 = MSE_2^*,
\end{equation}
\begin{equation}
T_c = 230.
\end{equation}
Here, $T_1$ and $T_4$ are the surface temperatures of the dayside and
nightside, respectively; $T_2$ and $T_3$ are the free-tropospheric
temperatures of the dayside and nightside, respectively; $T_c$ is the
cloud emission temperature; $S_0$ is the stellar flux at the
substellar point and $\frac{1}{2}$$S_0$ is the stellar flux impinging
on the dayside surface (averaged over the stellar zenith angle); 
$\alpha_p$ is the planetary albedo; $F_c$ is the
convective heat flux from the boundary layer to the free troposphere
in the dayside column; $F_a$ and $F_o$ represent atmospheric and
ocean heat transports from dayside to nightside, respectively; $F_d$
is the adiabatic heating due to dry descent in the nightside column;
$\varepsilon_2$ and $\varepsilon_3$ are the free tropospheric 
emissivities of the dayside and nightside, respectively; $f_c$ is the
effective cloud fraction; $MSE_1$ is the dayside surface moist static
energy; and $MSE_2^*$ is the dayside free-tropospheric saturation moist
static energy. Using the six Equations (2--7), we are able to determine 
solutions for the six dependent variables of the model: surface temperature 
($T_1$ \& $T_4$), free tropospheric temperature ($T_2$ \& $T_3$), atmospheric
heat transport ($F_a$), and dayside convective heat flux ($F_c$). This
solution is a function of the independent variables: the strength of
stellar flux ($S_0$), ocean heat transport ($F_o$), and model parameters.

Following the WTG approximation, the atmospheric energy transport
($F_a$) is forced to be strong enough to drive the dayside and
nightside free-tropospheric temperature together (Eq.~(6)). This
approximation allows the details of atmospheric dynamics to be avoided
but retains the net effect of atmospheric dynamics on the air
temperature. We assume that some of $F_a$, the total atmospheric
  heat transport from the dayside to the nightside, is deposited in
  the boundary layer ($F_d$) and the rest is deposited in the free
  troposphere ($F_a - F_d$). We take $F_d$ to be
 \begin{equation}
F_d =  k_1 F_a, 
\end{equation}
where $k_1$ is a constant parameter (see Table~1).

The surface and clouds are assumed to emit like blackbodies  
(i.e., $\varepsilon$\,$=$\,1).
The free troposphere is assumed to radiate as an equivalent 
gray body with emissivities of $\varepsilon_2$ and 
$\varepsilon_3$, determined by water vapor concentration  
\begin{equation}
\varepsilon_2,_3 = 1.0 - \exp(-\frac{\varkappa\,P_a}{g}\,q_2,_3) = 1.0 - \exp(-k_2\,q_2,_3), 
\end{equation}
where $q_2$ and $q_3$ are specific humidities of the dayside and
nightside respectively, $\varkappa$ is an absorption coefficient, $q$
is the specific concentration of water vapor, $P_a$ is the air
pressure, and $g$ is the acceleration due to gravity, and $k_2$ is a
constant coefficient with $k_2 \equiv \varkappa\,P_a/g$.  Equation
(10) follows from the gray gas approximation (e.g.,
\citet{Pierrehumbert2010}), and is similar to that employed in
\citet{Emanuel2002}.  We calculate the specific humidity by assuming
fixed relative humidities ($RH_2$ and $RH_3$, the ratio of water vapor
partial pressure to saturation water vapor partial pressure) and
calculating the saturation water vapor partial pressure using the
Clausius-Clapeyron relationship. By the Clausius-Clapeyron
relationship, saturation water vapor partial pressure is roughly
exponential in air temperature, increasing $\approx$7\% for each 1~K
in temperature at Earth-like temperatures \citep{Emanuel1994}. An
additional term in Eq.~(10) could be easily added to account for the
change of optical thickness due to atmospheric CO$_2$ absorption (for
example, see \citet{Emanuel2002}).

\begin{table}
\begin{center}
\caption{Model parameters used for the reference climate calculations.\label{tbl-1}}
\begin{tabular}{lll}
  \tableline\tableline
  Parameter & Description & Value \\
  \tableline
  $P_a$ & the depth of convection & 600 $hPa$ \\
  $T_c$ &  the cloud emission temperature & 230~$K$ \\
  $RH_1$ &  relative humidity of the dayside boundary layer & 90\% \\
  $RH_2$ &  relative humidity of the dayside free troposphere & 80\% \\
  $RH_3$ &  relative humidity of the nightside free troposphere & 30\% \\
  $k_1$ &  fraction of heat transport to the nightside  & 0.2 \\
   &  deposited in the boundary layer (Eq. (9)) & \\
  $k_2$ &  relates water vapor to the infrared opacity (Eq. (10)) & 1000 \\
  $k_3$ &  relates convection to the cloud fraction (Eq. (13)) & 0.08 \\
  \tableline
\end{tabular}
\end{center}
\end{table}

\begin{figure*}[]
\begin{center}
\vspace{0mm}
\begin{center}
\includegraphics[angle=0, width=34pc]{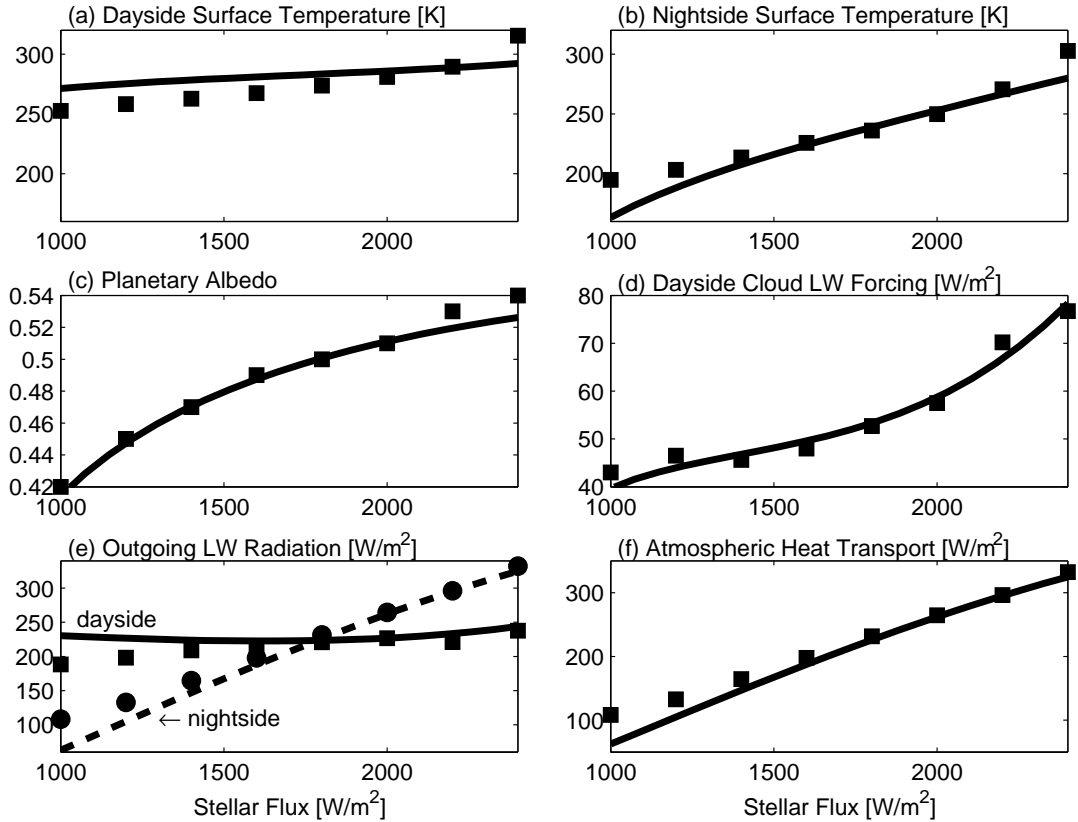}
\end{center}
\vspace{0mm}
\caption{Reference climate solution of the two-column model as a
  function of stellar flux ($S_0$) compared with output from the GCM
  CAM3. Lines represent the results of the two-column model, and
  squares and circles represent those of the GCM CAM3.  (a) dayside surface
  temperature ($T_1$); (b) nightside surface temperature ($T_4$); (c)
  planetary albedo ($\alpha_p$); (d) dayside cloud longwave forcing ($C_l$);
  (e) outgoing longwave radiation on the dayside ($OLR_1$) and on the
  nightside ($OLR_2$); and (f) atmospheric heat transport from dayside
  to nightside ($F_a$). Model parameters are listed in Table~1. 
  The squares and circles in (e) are the same as those in Fig. 4(b) for comparison.}
\label{fig5}
\end{center}
\end{figure*}

The convective heat flux ($F_c$) is determined
implicitly by requiring convective neutrality, which can be
approximated by requiring that the boundary layer moist static energy
equals the free-tropospheric saturation moist static energy
\citep{AbbotandTziperman2009}. The moist static energy
of the surface is
\begin{equation}
MSE_1 = C_p T_1 + L q_1,
\end{equation}
while the free-tropospheric saturation moist static 
energy is 
\begin{equation}
MSE_2^* = C_p T_2 + L q_2^* + g Z_a,
\end{equation}
where $C_p$ is the specific heat of the air at constant pressure
($C_p$\,=\,1005.7~$J\,kg^{-1}\,K^{-1}$), $L$ is the latent heat of evaporation
($L$\,=\,2.501$\times$10$^6$\,$J\,kg^{-1}$), $q_1$ is the surface specific humidity, 
$q_2^*$ is the free-tropospheric saturation specific humidity, $g$ is the
planetary gravitational acceleration, and
$Z_a$ is the height of convection, which we take to be a constant.

We set the relative humidity of the dayside free troposphere, $RH_2$,
to 80\% and of the nightside free troposphere, $RH_3$, to 30\%,
roughly the mean values we obtain from CAM3 simulations over a variety
of stellar fluxes. Similarly, we set the dayside 
boundary layer relative humidity, $RH_1$, to 90\%. 
We calculate the height of convection, $Z_a$, by
specifying the pressure of the free tropospheric layer, $P_a$, and
assuming a scale height of 5000~$m$. This scale height is smaller than
that of Earth, $\approx$7500~$m$, due to the fact that we use a larger
gravitational acceleration (13.7 versus 9.8\,$m\,s^{-2}$).


\begin{figure*}[]
\begin{center}
\vspace{0mm}
\begin{center}
\includegraphics[angle=0, width=34pc]{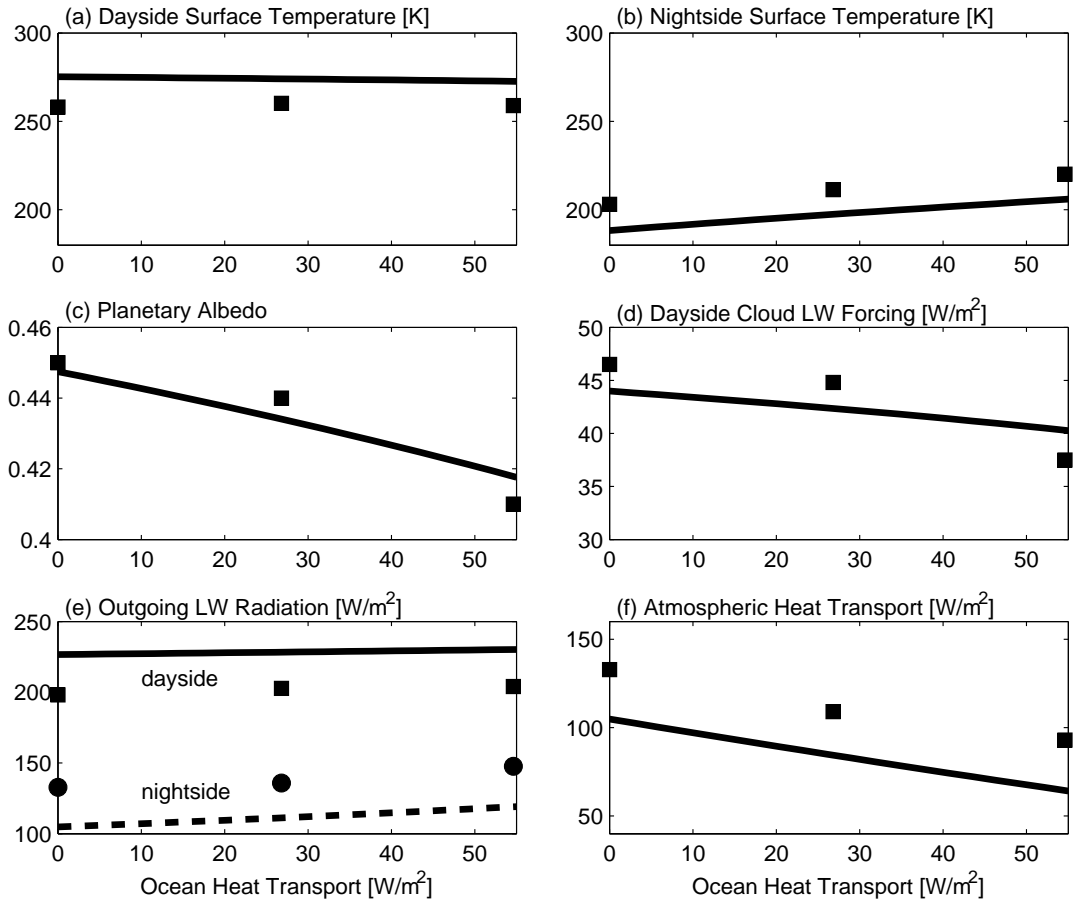}
\end{center}
\vspace{0mm}
\caption{As in Fig.~\ref{fig5}, but as a function of ocean heat
  transport from dayside to nightside ($F_o$). The stellar flux is
  1200\,W\,m$^{-2}$ in these calculations. Model parameters except for
  the parameter varied ($F_o$), are the same as those for the
  reference climate calculations (Table~1).  The moderate differences
  between the two-column model and GCM are due to the difference in
  the base state with $F_o$=0 (Fig.~\ref{fig5}), and can be removed by
  changing either the parameter $k_1$ or $k_2$ slightly.}
\label{fig6}
\end{center}
\end{figure*}

A simple convective cloud scheme is employed to describe the cloud
behavior. The effective cloud fraction ($f_c$) is related to the
convective heat flux ($F_c$) with an upper limit of 1.0 
\begin{equation}
f_c =  min(k_3 \ln(F_c+1.0), 1.0), 
\end{equation}
where $k_3$ is an adjustable parameter that we set to $k_3$\,=\,0.08
by tuning the model to roughly reproduce cloud albedos obtained in
CAM3 simulations. This logarithmic scheme 
for convective clouds was proposed by \citet{XuandKrueger1991} and used to 
  parameterize the cloud fraction in CAM3 \citep{Collinsetal2004} 
  and its recently released versions, CAM4 \citep{Nealeetal2010a} and CAM5 
  \citep{Nealeetal2010b}, and it is able to roughly simulate the convective 
  cloud coverage over the tropics of Earth (Hack et al. 2006).  
  We assume that the nightside column is
cloud-free because clouds there can have no albedo effect and 
have a small greenhouse effect because they are near the surface.

Cloud albedo depends on both cloud fraction and cloud microphysics.
For simplicity, detailed microphysical descriptions
\citep{Kitzmannetal2010, Zsometal2012} are not included in the model,
and we assume that the cloud albedo equals one where there are clouds. Using
this assumption, the planetary albedo ($\alpha_p$), determined by the
combined effect of surface and cloud albedos, is
 \begin{equation}
\alpha_p = 0.09 + f_c - 0.09 \times f_c, 
\end{equation}
where 0.09 is the albedo of seawater averaged over the dayside 
(including the dependence of albedo on the stellar zenith angle). 
If there are no clouds, the
planetary albedo is equal to the seawater albedo. We do not include
the contribution of sea ice and snow to the planetary albedo, which would
only be relevant near the outer edge of the habitable zone when
significant amounts of sea ice and snow spread to the dayside. 

We specify the cloud emission temperature ($T_c$) to be a constant in
accordance with the FAT hypothesis. We set $T_c$\,=\,230~K, which is
the value suggested by our GCM simulations (see Fig.~3) and is 12~K 
higher than that of tropical anvil clouds of
Earth \citep[$\approx$218~K;][]{HouzeandBetts1981}.

We neglect a variety of physical processes in the model. For example,
we do not consider shortwave absorption by water vapor and clouds, 
feedbacks due to changes of convection depth and
relative humidity, the detailed interaction between
atmospheric and oceanic circulations, and the momentum and moisture
budgets. Despite these simplifications, the low-order two-column model
is able to capture many of the fundamental features of the GCM
simulations.


\section{Reference Climate and Model Sensitivity}

In this section, we first show that it is possible to tune the
two-column model to reproduce the results of the GCM CAM3
simulations. Next, we investigate the sensitivity of the model to
increases in ocean heat transport, cloud radiative forcing, and 
atmospheric emissivity.

\subsection{Reference Climate}

We plot model variables at steady state for the reference climate as a
function of stellar flux in Fig.~5 (see Table~1 for model
parameters). The ocean heat transport is set to zero in these
calculations. We choose the values of $P_a$, $T_c$, $RH_1$, $RH_2$,
and $RH_3$ by averaging equivalent variables in the CAM3 GCM
simulations. We then tune the two-column model to CAM3 by adjusting
only three parameters: $k_1$, $k_2$, and $k_3$.  When the two-column
model is tuned to CAM3 in this way, it is able to capture the
fundamental features of the climate, including surface temperature,
planetary albedo, cloud longwave forcing, thermal
emission flux, and atmospheric heat transport.

Both the dayside and nightside surface temperatures increase as the
stellar flux increases, and their magnitudes are close to CAM3's. The
planetary albedo of the two-column model behaves very similarly to
that of CAM3, which shows that our simple convective cloud scheme is
capable of capturing the bulk effects of cloud albedo on the
climate. The planetary albedo of tidally locked terrestrial planets
(between 0.42 and 0.54) is much higher than that of Earth
($\approx$0.3). This results from a high cloud fraction near the
substellar point, where the stellar flux is highest (YCA13). The two-column
model also has very similar dayside cloud longwave forcing to that of
CAM3 (between 40 and 80\,W\,m$^{-2}$).

As the stellar flux increases, both models predict that the dayside
thermal emission flux is nearly constant whereas the nightside thermal
emission increases rapidly (Fig.~5e). When the stellar flux is higher
than $\approx$1800\,W\,m$^{-2}$, the thermal emission flux on the
nightside exceeds that on the dayside. The strength of atmospheric
heat transport from dayside to nightside is quite similar between the
two-column model and CAM3 (Fig.~5f), confirming the accuracy of the
WTG approximation.


\begin{figure}[h!]
\vspace{0mm}
\begin{center}
\includegraphics[angle=0, width=17pc]{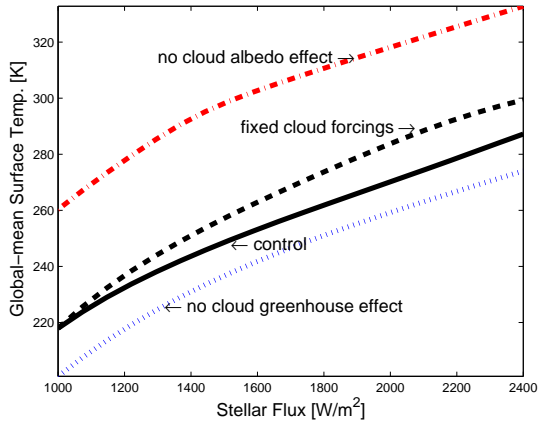}
\end{center}
\vspace{0mm}
\caption{Sensitivity of global-mean surface temperature ($\frac{T_1+T_4}{2}$) to 
cloud albedo and cloud longwave forcing, as a function of stellar flux ($S_0$). 
Solid line: with both effects; dashed-dotted line: no cloud albedo effect ($\alpha_p$\,=\,0.09); dotted line: no 
cloud greenhouse effect ($C_l$\,=\,0); and dashed line: planetary albedo 
and cloud greenhouse effect are fixed to those at a stellar flux of  
1000\,W\,m$^{-2}$  ($\alpha_p$\,=\,0.415, 
$C_l$\,=\,40\,W\,m$^{-2}$; i.e., no cloud feedback). 
Other model parameters are the same as those for the reference climate 
calculations (Table~1).}
\label{fig7}
\end{figure}

Figure~5 shows that as the stellar flux increases, the nightside
surface temperature increases much faster than the dayside surface
temperature. The main reason for this is that the atmosphere
efficiently transports energy absorbed on the dayside to the nightside
(Fig.~5f).

These comparisons between the two-column model and CAM3 demonstrate
that despite of the simplicity of our assumptions, the two-column
model is able, when appropriately tuned, to reproduce the main climate
features of tidally locked terrestrial planets as simulated by a
GCM. This indicates that the climate of tidally locked planets is well
described by the handful of essential physical processes contained in
the two-column model.


\subsection{The Effect of Ocean Heat Transport}


Here we investigate how the climate responds to imposed ocean heat
transport from dayside to nightside. To investigate this we performed
a series of calculations in which the ocean heat transport is varied
while the stellar flux is fixed to a constant (Fig.~6). As the ocean
heat transport is increased, the dayside surface temperature decreases
slightly while the nightside surface temperature increases
significantly, so that the global-mean surface temperature increases,
and the surface temperature and thermal emission contrasts between
dayside and nightside decrease. Both the planetary albedo and the
cloud longwave forcing decrease because increasing ocean heat
transport decreases the day--night surface temperature contrast,
weakening the convergence and convection on the dayside. The
atmospheric heat transport decreases in response to the increase in
ocean heat transport, such that the total heat transport shows a small
increase. This compensation between the atmosphere and ocean is
consistent with the results of simulations with a fully coupled 3D
atmosphere-ocean circulation model \citep{HuandYang2014} and is
similar to the behavior that meridional atmospheric and ocean heat
transports on Earth are believed to exhibit \citep{Stone1978,
  FarnetiandVallis2013}.

We tuned the two-column model to the GCM over a range of stellar
fluxes, and it deviates somewhat from the GCM when the ocean heat
transport is zero at the particular stellar flux for which we varied
the ocean heat transport in the GCM (Fig.~6). Nevertheless, the
two-column model does a good job of reproducing the trends of the GCM
as the ocean heat transport is increased. Moreover, by slightly changing 
$k_1$ or $k_2$ we can greatly improve the match between the GCM and
the two-column model when the ocean heat transport is varied at this
particular stellar flux (not shown).

The fact that the two-column model can roughly reproduce the behavior
of the GCM when ocean heat transport is varied, even though it was
tuned to match the GCM when stellar flux is varied, suggests that the
approximations we have made capture the relevant physics, rather than
the model simply being cleverly tuned.  For example, the fact that the
response of atmospheric heat transport to changes in ocean heat
transport in the two-column model is very similar to that in the GCM
emphasizes the validity of the WTG approximation. Furthermore, the
similar changes in planetary albedo and cloud longwave forcing between
the models suggests that the convective scheme and FAT hypothesis are
faithful representations of the more complete physics contained in the
GCM.


\begin{figure*}[]
\begin{center}
\vspace{0mm}
\begin{center}
\includegraphics[angle=0, width=34pc]{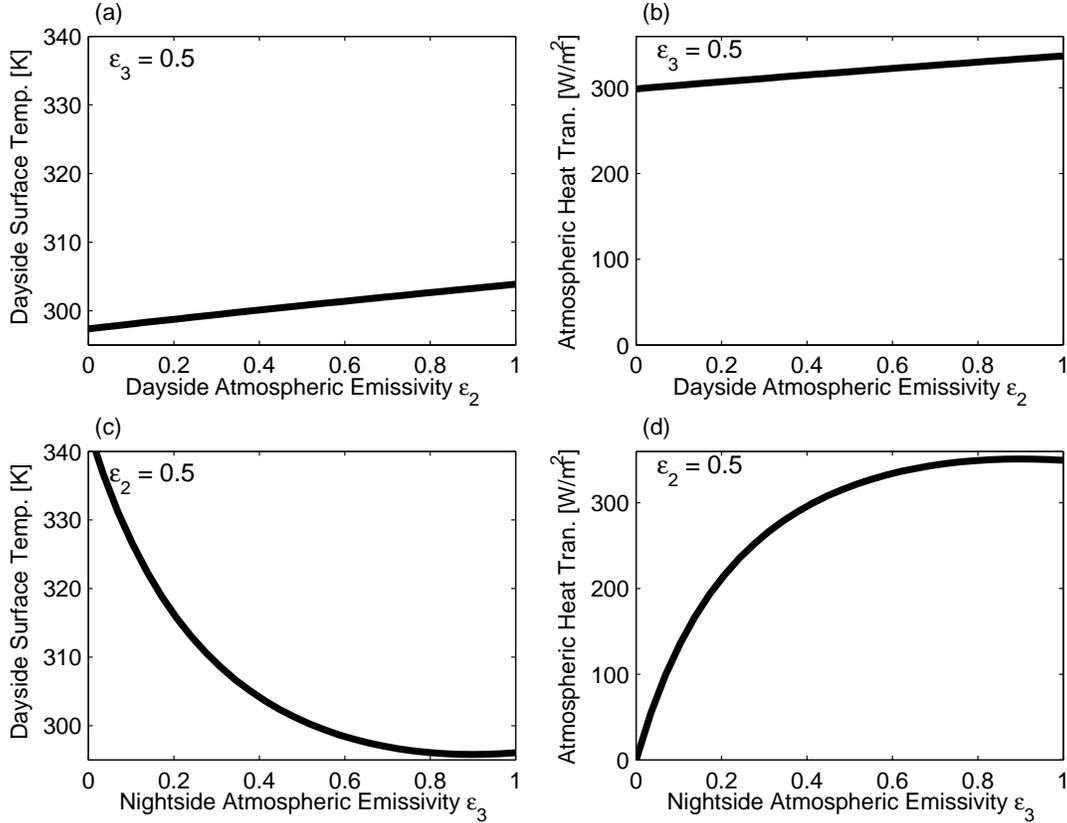}
\end{center}
\vspace{0mm}
\caption{Sensitivity of the climate to atmospheric emissivity of dayside 
($\varepsilon_2$) and nightside ($\varepsilon_3$). 
Left panels: dayside surface temperature ($T_1$); 
right panels: atmospheric heat transport from dayside to nightside ($F_a$).
(a) \& (b): Varying $\varepsilon_2$ but fixed $\varepsilon_3$ to 0.5; 
(c) \& (d): varying $\varepsilon_3$ but fixed $\varepsilon_2$ to 0.5. 
The stellar flux is 2400\,W\,m$^{-2}$. 
Planetary albedo and cloud longwave forcing are fixed 
($\alpha_p$\,=\,0.415, $C_l$\,=\,40\,W\,m$^{-2}$). 
Other model parameters are the same as those for the reference climate calculations.}
\label{fig8}
\end{center}
\end{figure*}


\subsection{The Importance of Cloud Albedo Effect}

Clouds have significant effects on the climate of tidally locked
planets (YCA13). In order to further establish the importance of
clouds, we performed calculations with either the cloud albedo or cloud
longwave forcing set to zero (Fig.~7). For the no cloud albedo 
effect experiment, Eq.~(14) is not used and the planetary albedo 
is set to equal to the seawater albedo (0.09). For the no cloud 
longwave effect experiment, Eq.~(13) is not used and the 
cloud longwave forcing is set to zero. 



Comparisons with the reference climate experiments show that the
global-mean surface temperature decreases by $\approx$15~K  
for the no cloud longwave effect experiments while it increases 
by $\approx$40~K if the cloud albedo effect is turned off.  
This indicates that the cloud albedo
cooling effect is much stronger than the cloud greenhouse warming
effect. The net effect of the clouds is therefore to dramatically cool
the planet, delaying the onset of a runaway greenhouse, 
as addressed in YCA13.

We also performed a series of calculations with a fixed planetary
albedo of 0.415 and a fixed cloud longwave forcing of 40\,W\,m$^{-2}$,
which are their values when the stellar flux is equal to
1000\,W\,m$^{-2}$. These calculations investigate the model with
clouds, but without a cloud feedback (changes in cloud forcing as
a function of climate). The dashed line of Fig.~7 shows that the
surface temperature increases by $\approx$15~K on the dayside, 
relative to the interactive cloud case, when the stellar flux is 2400\,W\,m$^{-2}$. 
The cloud albedo feedback therefore has a significant cooling effect on the planet.


\subsection{The Importance of the Nightside Radiator Fin}


Atmospheric emissivity affects the strength of the greenhouse effect
and the efficiency of infrared energy loss to space. We perform a set
of experiments to test the sensitivity of the climate to atmospheric
emissivity. In these experiments, Eq.~(10) is not used, and instead
atmospheric emissivity is set to a series of values from zero to
one. The cloud radiative forcing is fixed in order to focus on the
role of atmospheric emissivity.

Figure~8a shows the dayside surface temperature  
as a function of the dayside atmospheric emissivity ($\varepsilon_2$). 
In these experiments, the nightside atmospheric emissivity ($\varepsilon_3$) 
is fixed to 0.5. As $\varepsilon_2$ is increased, 
the dayside atmospheric greenhouse effect increases, and consequently
the surface temperature increases. The magnitude of the increase, however, 
is small even when $\varepsilon_2$ is increased from zero to one. 
This is due to an increase in atmospheric heat export to 
the nightside (Fig.~8b), which warms the nightside and weakens the increase in 
the dayside surface temperature.

Figure~8c shows the sensitivity of the climate to varying the
nightside atmospheric emissivity ($\varepsilon_3$). In these
experiments, the dayside atmospheric emissivity ($\varepsilon_2$) is
fixed to 0.5. As $\varepsilon_3$ is increased, the dayside surface
temperature decreases strongly, although the nightside and global-mean
surface temperatures increase (not shown).  As $\varepsilon_3$ is
increased from zero to 1.0, the dayside surface temperature decreases
by $\approx$45~K. As $\varepsilon_3$ is increased, the efficiency of the
atmospheric infrared energy loss to space increases, which requires a great
increase in atmospheric heat transport from dayside to nightside
(Fig.~8d), therefore cooling the dayside surface.


\begin{figure}[h!]
\vspace{0mm}
\begin{center}
\includegraphics[angle=0, width=17pc]{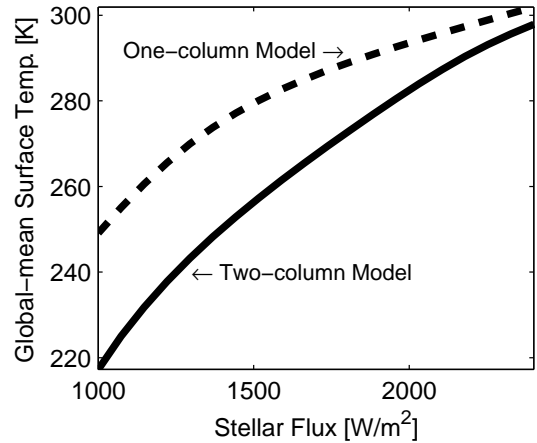}
\end{center}
\vspace{0mm}
\caption{The effect of the radiator fin on the global-mean surface temperature. 
The one-column model (without the radiator fin) has the same parameters 
of the two-column model (including the radiator fin), including planetary albedo 
($\alpha_p$\,=\,0.415), cloud longwave forcing ($C_l$\,=\,40\,W\,m$^{-2}$), 
relative humidity, and atmospheric emissivity (Table~1).}
\label{fig9}
\end{figure}

We can draw an important physical lesson 
from these experiments: the nightside acts like a
``radiator fin'', similar to the subtropics of Earth
\citep{Pierrehumbert1995}, only more pronounced. All the stellar
energy is deposited on the dayside and the water vapor and cloud
greenhouse effects prevent the atmosphere from getting rid of all this
energy locally. Much of the energy is instead transported by the atmosphere
 to the nightside where the air is dry, there is a strong
temperature inversion, and the cloud greenhouse effect is negligible, so that
infrared energy is easily emitted to space (as it would be by a ``radiator
fin''). The nightside atmosphere therefore has a stabilizing effect on
the climate of the planet.

The radiator-fin effect cannot be captured in a single column model,
even a 1D radiative-convective model with an extremely sophisticated
radiative transfer scheme, because it requires horizontal heat
transport to connect the moist and dry regions. In order to
demonstrate this, we construct a one-column model with the same
parameters as the two-column model.  A comparison of the surface
temperature as a function of the stellar flux in the one-column
and two-column models is shown in Fig.~9. We find that the surface
temperature of the one-column model is $\approx$15--30~K higher than
the two-column model because the one-column model is not able to
capture the effect of the radiator fin. The radiator-fin effect
becomes weak at high stellar fluxes because the atmospheric
emissivities of both the dayside and nightside approach one. This is a
limitation of the simple radiative scheme in our model. Further
investigation of the effect of the radiator fin on the runaway
greenhouse will require a 2D or 3D model with a more sophisticated 
radiative scheme.


\section{Day--night Thermal Emission Contrast}

YCA13 proposed that cloud behavior on tidally locked terrestrial
planets near the inner edge of the habitable zone would cause higher
infrared thermal emission from the nightside than the dayside. This
would lead to a reversal in the thermal phase curve, which is a
qualitatively different signal than would normally be expected. In
this section, we consider the difference in thermal emission flux
between dayside and nightside in the two-column model, and show how
basic physical parameters may affect this signal.

In a steady state, the dayside column heat budget 
can be obtained by summing Eqs. (2) and (3),
\begin{equation}
\frac{1}{2}S_0(1-\alpha_p) - F_a - F_o = OLR_1.
\end{equation}
$OLR_1$ is the thermal emission flux from the dayside, with 
\begin{equation}
OLR_1 = (1-\varepsilon_2) \sigma T_1^4 + 
\varepsilon_2 \sigma T_2^4 - C_l, 
\end{equation}
where the sum of the first and second terms of the right hand side is the 
outgoing infrared emission in clear-sky conditions, and the third term 
is the reduction of the outgoing infrared emission due to the presence of clouds with 
\begin{equation}
C_l = f_c ((1-\varepsilon_2) \sigma T_1^4 + 
\varepsilon_2 \sigma T_2^4) - f_c \sigma T_c^4. 
\end{equation}

Figure~10a shows the three terms of Eq.~(16) in the reference climate
experiments. Surface emission reaching the top of the atmosphere
decreases greatly with increasing stellar flux although the surface
temperature increases. This is due to the increase of atmospheric
absorption. Water vapor concentration increases rapidly with air
temperature because of the strong temperature dependence of saturation
water vapor pressure following the Clausius--Clapeyron relation,
reducing the thermal emission to space. Emission by the atmosphere
increases with increases in stellar flux due to increases of the air
temperature ($T_2$) and atmospheric emissivity ($\varepsilon_2$).  The
cloud longwave effect decreases with increasing stellar flux, also
reducing the thermal emission to space. The cloud effect is comparable
to that of water vapor. Due to the combined effect of water vapor and
clouds, the dayside $OLR$ curve as a function of the stellar flux is
very flat.  Figure 10a also shows that if the stellar flux is low, the
optical depth of the atmosphere is small, and $OLR_1$ is close to the
unmodified upward radiation from the surface. If the stellar flux is
high, the surface upward emission is blocked by the optically thick
atmosphere, and the outgoing thermal emission is primarily determined
by the upper tropospheric temperature.

\begin{figure*}[]
\begin{center}
\vspace{0mm}
\begin{center}
\includegraphics[angle=0, width=34pc]{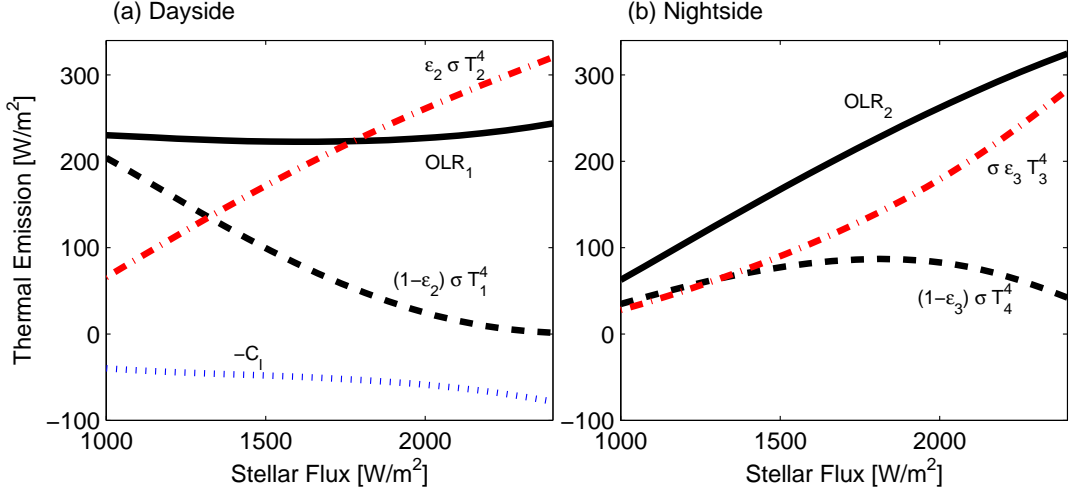}
\end{center}
\vspace{0mm}
\caption{Dayside (a) and nightside (b) thermal emission fluxes 
as a function of the stellar flux ($S_0$) between 1000~and~2400~W\,m$^{-2}$. 
Solid line: thermal emission flux to space, which has to 
be equal to the sum of other terms (see Eqs.~(16) \& (19)); dashed line: surface upward emission 
minus the absorption by water vapor; dashed--dotted line: atmospheric 
reemission; and dotted line: the effect of clouds. Model parameters 
are the same as those for the reference climate calculations (Table~1).}
\label{fig10}
\end{center}
\end{figure*}

For the thermal emission flux on the dry nightside, water vapor has a
small effect, and low-level clouds there have nearly zero effect. The
nightside column heat budget can be obtained by summing Eqs. (4) and
(5),
\begin{equation}
F_a + F_o = OLR_2,
\end{equation}
where $OLR_2$ is the thermal emission flux to space: 
\begin{equation}
OLR_2 = (1-\varepsilon_3) \sigma T_4^4 + \varepsilon_3 \sigma T_3^4.
\end{equation} 
This means that the thermal emission flux on the nightside is equal to
the sum of heat transports from the dayside by the atmosphere and
ocean. Figure~10b shows the two terms of $OLR_2$ as a function of
stellar flux. As the stellar flux increases, surface emission reaching
the top of the atmosphere first increases due to the increase in
surface temperature, and then decreases due to the increase of water
vapor absorption.  Atmospheric emission increases with increasing
stellar flux due to increases of air temperature and emissivity. The
net effect of the two terms is a strong increase in the thermal
emission flux with increasing stellar flux.

If we assume a saturated atmosphere on the dayside, there will exist
an upper limit for $OLR_1$ at which the atmosphere becomes optically
thick at all infrared wavelengths \citep{Kastingetal1993}. The limit
is $\approx$280\,W\,m$^{-2}$ for planets with pure water atmospheres,
Earth's gravity, and no clouds \citep{Goldblattetal2013}. If clouds
are further considered, the limit will be lower.  On the nightside,
thermal infrared energy is more easily emitted to space through the
radiator fin.

A remaining question is: What is the critical stellar flux ($S_c$) at
which the dayside outgoing thermal emission flux becomes equal to or
less than that of the nightside, i.e., $OLR_1$$\le$$OLR_2$? If $OLR_1$
is higher than $OLR_2$, the phase curve maximum should occur when the
dayside of the planet is facing the observer. If $OLR_1$ is lower than
$OLR_2$, the phase curve maximum should occur when the nightside is
facing the observer.  This striking, qualitative reversal in the
thermal phase curve would be strong evidence for the Earth-like cloud
behavior, if observed. Because of this, it is important to understand
which model parameters exert the strongest control on $S_c$.

\begin{figure*}[]
\begin{center}
\vspace{0mm}
\begin{center}
\includegraphics[angle=0, width=34pc]{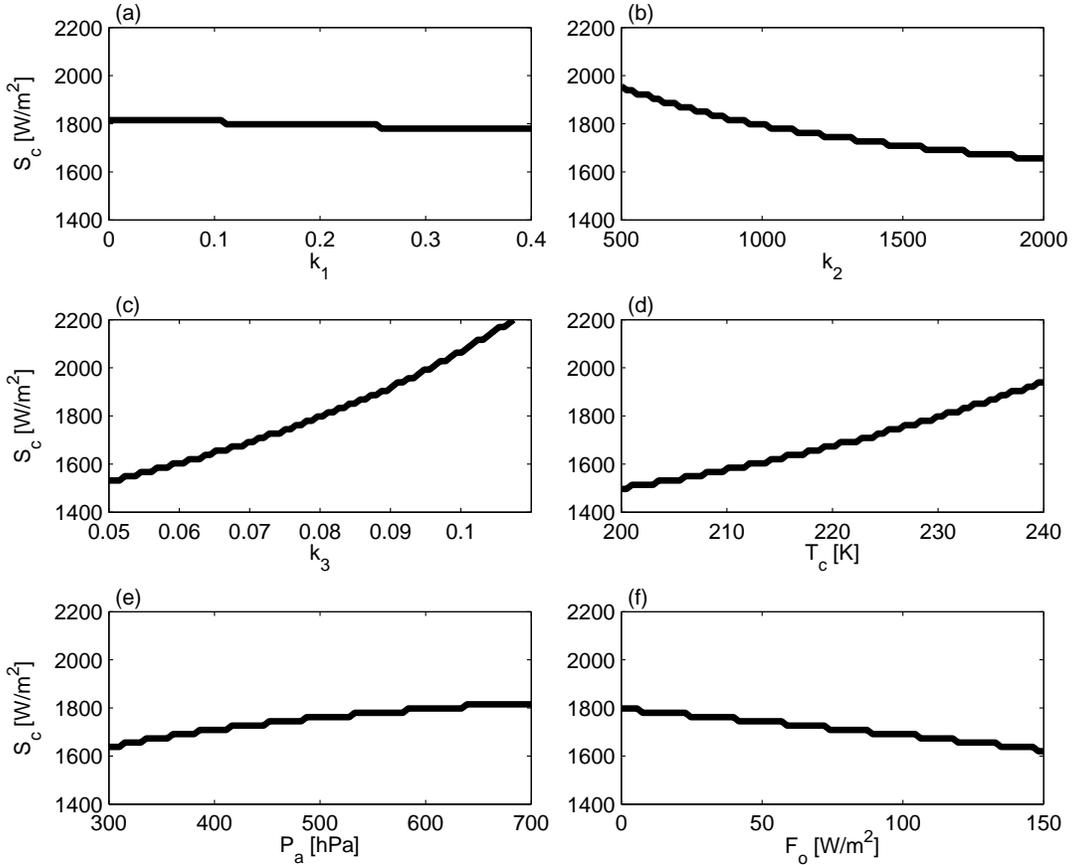}
\end{center}
\vspace{0mm}
\caption{The critical stellar flux ($S_c$) at which the thermal emission 
flux of the dayside becomes equal to or less than that of the nightside 
($OLR_1$$\le$$OLR_2$), as a function of: (a) $k_1$, 
a parameter in Eq.~(9), representing 
the strength of adiabatic heating on the nightside surface due to 
large-scale dry descent; (b) $k_2$, a parameter in Eq.~(10), 
representing the strength of water vapor absorption; (c) $k_3$, 
a parameter in Eq.~(13), representing the increase in cloud fraction 
when the convective heat flux increases; (d) $T_c$, the cloud emission 
temperature; (e) $P_a$, the depth of convection; and (f) $F_o$, 
the ocean heat transport. Model parameters, except for the parameter 
varied, are the same as those for the reference climate calculations (Table~1).}
\label{fig11}
\end{center}
\end{figure*}

In Fig.~11, we show how $S_c$ changes as we vary $k_1$, $k_2$, $k_3$,
$T_c$, $P_a$, and $F_o$, which are the most important independent
parameters of the model. Here, $k_1$ represents the strength of
adiabatic warming on the nightside surface due to large-scale dry
descent (Eq. (9)); $k_2$ is a coefficient relating water vapor to the
infrared opacity (Eq. (10)); $k_3$ is a constant that relates the
strength of convection to the cloud fraction (Eq. (13)); $T_c$ is the
cloud emission temperature; $P_a$ is the depth of convection; $F_o$ is
the strength of ocean heat transport.

Changing $k_1$, $P_a$, and $F_o$ does not influence the value of $S_c$
very much (Fig.~11).  In contrast, $k_2$, $k_3$, and $T_c$ strongly
influence $S_c$. Increasing $k_2$ increases the optical thickness due
to water vapor. Since there is more water vapor on the dayside, this
preferentially decreases $OLR_1$, and therefore decreases $S_c$. Although $k_2$
is a necessary parameter in the two-column model, models with a
more sophisticated radiative transfer scheme should agree on water
vapor emissivity, and therefore the processes represented by $k_2$ are
unlikely to be a major source of uncertainty in $S_c$.

The effects of $k_3$ and $T_c$ on $S_c$ are more likely to be relevant
for future studies. These parameters represent cloud processes, which
can differ significantly between complicated climate
models. Increasing the cloud emission temperature ($T_c$) decreases
the strength of cloud longwave forcing, increasing $OLR_1$ and
therefore increasing $S_c$.  Increasing $k_3$ causes an increase in
the cloud fraction for a given convective heat flux, which increases
both the cloud albedo and the cloud longwave forcing.  Because the
cloud albedo cooling effect is always stronger than the cloud
greenhouse warming effect, water vapor amount and consequently atmospheric
emissivity decrease. Since this primarily increases the $OLR$ on the dayside,
increasing $k_3$ causes $S_c$ to increase. Cloud behavior will
therefore be critical for determining the critical stellar flux
($S_c$), and a measurement of $S_c$ will help us distinguish between
different cloud models.

Finally we note that other factors that we have not explicitly
  included in the two-column model, such as atmospheric CO$_2$, can
  influence the day-night emission contrast. For example, simulations
  in CAM3 show that increasing the CO$_2$ concentration strengthens
  the day-night thermal emission contrast because it weakens the
  nightside inversion and leads to a stronger water vapor feedback on
  the warmer dayside (not shown). This can change the critical stellar
  flux by $\approx$400\,W\,m$^{-2}$ when the CO$_2$ concentration 
  is increased from 0 to 0.1 bar, so that large uncertainties in the CO$_2$ of an 
  observed planet would make it harder to fit a cloud model to the observed
  thermal emission contrast.


\section{Conclusions}

We constructed a low-order two-column model to simulate the climate of
tidally locked terrestrial planets near the inner edge of the
habitable zone of M-dwarf stars.  This model incorporates the
weak-temperature-gradient approximation to calculate horizontal
atmospheric heat transport, the fixed anvil temperature hypothesis to
calculate the cloud longwave effect, and a simple convective cloud
scheme to calculate the cloud albedo. Our main findings are as
follows:
 
(1) The relatively few parameters of the two-column model can be tuned
to reproduce the basic behavior of a very complicated global climate
model and respond faithfully to variations in both the stellar flux
and ocean heat transport. This suggests that the physical processes
built into the two-column model are the most important ones for
determining the climate of tidally-locked terrestrial planets.

(2) The two-column model clearly illustrates the importance of
dayside clouds and a nightside ``radiator fin'' for determining the
climate of tidally locked planets. Dayside clouds increase the
planetary albedo, which cools the planet and delays the onset of a
runaway greenhouse. The radiator fin, primarily maintained by the
dryness of the atmosphere and the temperature inversion on the
nightside, allows the planet to easily lose energy to space through
infrared thermal emission, which also cools the
planet. Atmospheric dynamics are crucial for both of these effects, so
they cannot be calculated by 1D radiative-convective models. This
means that 1D radiative-convective models will tend to produce overly
conservative estimates of the inner edge of the habitable zone.

(3) Observations of the day--night thermal emission contrast will be
critical for deciphering the climate of tidally locked terrestrial
planets. As the stellar flux increases, nightside emission increases
faster than dayside emission, so that at a critical stellar flux there
will be a reversal in the day--night thermal emission contrast, with
higher emission from the nightside. Sensitivity experiments in the
two-column model show that cloud variables are most likely to cause
differences in the critical stellar flux between different
models. This means that future observations of the day--night thermal
emission contrast will be useful for distinguishing between cloud
models.



\acknowledgments \textbf{Acknowledgments:} We are grateful to
Daniel~D.\,B.~Koll, Feng~Ding, Yuwei Wang, and Feng Tian for helpful
discussions and comments.  D.S.A. acknowledges support from an Alfred
P. Sloan Research Fellowship. We also thank two anonymous reviewers
for their constructive comments.
 



\end{document}